# Gaussian Beam Collimation via Transformation Media

**Yuancheng Fan,**[*] **Jin Han, and Hongqiang Li**

*Department of Physics, Tongji University, Shanghai 200092, China*

[*]*Corresponding Author: boyle429@126.com*

**Abstract:** Conventional method for Gaussian beam collimation is to expand the beam waist to achieve a smaller divergence angle. Recent transformation optics offers an unconventional path to control the electromagnetic field. In this paper we show that a Gaussian beam can be converted into a plane wave-like beam by transformation media within the scale of several wavelengths. Design details and full-wave simulation results via finite element method are provided.

## 1. Introduction

In optical physics Gaussian beam refers to a beam of electromagnetic waves whose transverse electric field and intensity follow Gaussian distribution in space. Most directional sources, such as lasers in optical region and horn antennas in microwave region, emit electromagnetic (EM) waves in a form of Gaussian beam. Directional sources are useful in myriad applications; one direct benefit is to transmit or detect electromagnetic (EM) signals over long distance. One most commonly used method to collimate a Gaussian beam is to expand the beam waist through the two-lens confocal system[1, 2]. Based on the form-invariant coordinate transformation of Maxwell's equation, Pendry and Leonhardt independently proposed the concept of transformation media for EM waves, paving a new way to design EM devices with specific functionalities[3, 4] and unprecedented abilities to manipulate EM waves, such as invisible cloak[3], super-lenses and hyper-lenses[5-9], EM wave rotator[10, 11], waveform transformer[12] and electromagnetic wormhole[13] etc. It has been shown experimentally that transformation media is practically attainable with the development of metamaterials [14-18].

It is noted that finite embedded coordinate transformation technique[19] recently proposed by M. Rahm et al. shall bring less singular material parameters to transformation media than the transformation from one point to a volume[14, 20]. The method was applied successfully in [21] and [22] to collimate cylindrical waves. Here we advanced the work to the widely used Gaussian beam specially. Design details and full-wave simulations using a finite element method are provided.

## 2. Gaussian beam collimation

We consider a Gaussian beam propagating along the $x$ direction. One of its wave components shall be expressed by

$$A(r,x) = A_0 \frac{\omega_0}{\omega(x)} \exp\left[-\frac{r^2}{\omega^2(x)}\right] \exp\left[-i\left[k\left(x+\frac{r^2}{2R(x)}\right) - arc\tan\frac{x}{f}\right]\right] \quad (1)$$

where $r = \sqrt{y^2 + z^2}$ is radial distance of a point from center axis of the beam. $A_0$ is a constant denoting the beam amplitude, $R(x)$ and $\omega(x)$ are measures of beam width and wave-front radius of curvature respectively shown as solid and dashed curves in Fig.1. $\omega_0 = \omega(x)|_{x=0}$ is known as beam waist at $x = 0$ plane, $k = 2\pi/\lambda$ is wave-vector in free-space with $\lambda$ being wavelength, and $f$ is a real number being Rayleigh range. The divergence angle of the monochromatic beam operating at wavelength $\lambda$ is measured by $\theta_0 \simeq \lambda/\pi\omega_0$, which means that a smaller beam waist will bring more divergence to the Gaussian beam. A conventional method for Gaussian beam collimation is to expand beam with a two-lens confocal system. The first lens is to focus a monochromatic beam at confocal point, and the second lens with a focal length $F$ is to expand and collimate the beam so that the system finally transforms the beam into another Gaussian beam with a smaller divergence angle $\theta = \dfrac{\pi\omega_0^2}{F\lambda}\theta_0$.

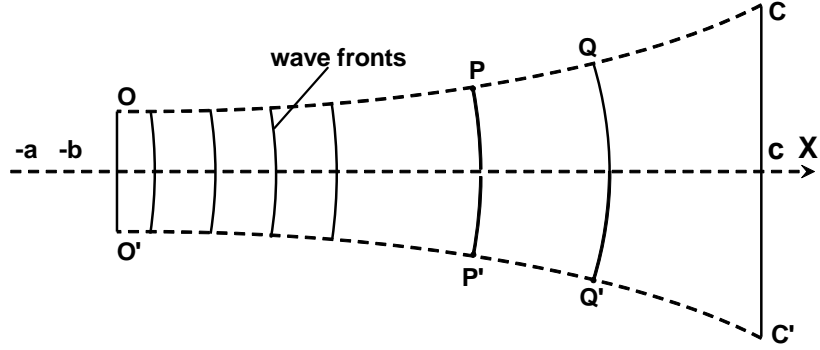

Fig. 1 Schematic configuration of Gaussian beam and space domain for transformation. Wave-front curves $P'P$ and $Q'Q$ segment the transformed region into region $I$, $II$ and $III$.

The above-mentioned conventional method for beam collimation starts from the view of geometric optics, while transformation media provide us a new vision to collimate a beam within several wavelengths in scale. We now present how to realize Gaussian beam collimation by coordinate transformation. For simplicity, we consider the transformation in two dimensional (2D) spaces provided that a Gaussian beam is symmetric to $x$ axis in spatial field distributions following EQ. (1). It is well known that coordination transformation staring from a line segment [20] or an area [19] usually requires smaller and less singular material parameters than the one directly transformed from a point[14]. Here we deploy a two-step transformation inspired by the embedded transformation scheme [19], which shall gives rise to smoothly varying distribution of material parameters in space.

Specifically we adopt two wave-fronts of Gaussian beam $QQ'$ and $PP'$ at $x=-b$ and at $x=-a$, so that the space domain (see Fig.1) is divided into three trapezoidal-like regions $I$ (between $OO'$ and $PP'$), $II$ (between $PP'$ and $QQ'$), and $III$ (between $QQ'$ and $CC'$). The two wave-fronts have radii $R_1$ and $R_2$ respectively. We deploy a two-step scheme to perform coordination transformation from original space $(x, y)$ to real space $(x', y')$. The first step is to shift trapezoidal-like region $II$ to superpose region III by the mapping function

$$\frac{r-R_2}{D(y)-R_2} = \frac{r'-D(y)}{\dfrac{(a+c)r}{x}-D(y)} \qquad (x,y) \in II \qquad (2)$$

where $D(y) = \sqrt{\left(\sqrt{R_1^2 - y^2} - b + a\right)^2 + y^2}$. And the second step is to extend region $I$ into a larger area to precisely superpose region $I$ and the void region $II$. In cylindrical

coordinates, the corresponding mapping function of the transformation in second step follows the form of

$$\frac{r - \dfrac{ar}{x+a}}{R_2 - \dfrac{ar}{x+a}} = \frac{r' - \dfrac{ar}{x+a}}{D(y) - \dfrac{ar}{x+a}} \qquad (x, y) \in I \qquad (3)$$

And in Cartesian coordinates, the relations in EQ. (2) and EQ. (3) shall be expressed as follows

$$\begin{pmatrix} x' \\ y' \\ z' \end{pmatrix} = \begin{pmatrix} \dfrac{x(x+a)D(y) - axr}{R_2(x+a) - ar} \\ \left(D(y) - \dfrac{ar}{x+a}\right)\dfrac{y(x+a) - ay}{R_2(x+a) - ar} + \dfrac{ay}{x+a} \\ z \end{pmatrix} \qquad (4)$$

for $(x', y') \in I + II$ and

$$\begin{pmatrix} x' \\ y' \\ z' \end{pmatrix} = \begin{pmatrix} \dfrac{r - R_2}{D(y) - R_2}\left(\dfrac{(a+c)(x+a)}{x} - \dfrac{x+a}{r}D(y)\right) + \dfrac{x+a}{r}D(y) - a \\ \dfrac{r - R_2}{D(y) - R_2}\left(\dfrac{(a+c)y}{x} - \dfrac{y}{r}D(y)\right) + \dfrac{y}{r}D(y) \\ z \end{pmatrix} \qquad (5)$$

for $(x', y') \in III$ respectively.

By utilizing the method of transformation optics presented in Ref [3, 19], we shall obtain the transformation media parameters through $\overline{\varepsilon}'(r') = \Lambda\Lambda^T/\det\Lambda = \overline{\mu}'(r')$ where $\Lambda$ is Jacobian matrix with $\Lambda_\alpha^{\alpha'} = \partial r'^{\alpha'}/\partial r^\alpha$ as matrix elements. In terms of EQ. (4) and EQ. (5) we shall derive $\overline{\Lambda}$ for coordinate transformation

$$\overline{\Lambda} = \begin{pmatrix} \dfrac{\partial x'}{\partial x} & \dfrac{\partial x'}{\partial y} & 0 \\ \dfrac{\partial y'}{\partial x} & \dfrac{\partial y'}{\partial y} & 0 \\ 0 & 0 & 1 \end{pmatrix} = \begin{pmatrix} a_{11} & a_{12} & 0 \\ a_{21} & a_{22} & 0 \\ 0 & 0 & 1 \end{pmatrix} \qquad (6)$$

Under the assumption about $\bar{\varepsilon} = \bar{\mu} = \bar{I}$ in original space, we get the relative permittivity $\bar{\varepsilon}'$ and permeability $\bar{\mu}'$ in transformed real space

$$\bar{\varepsilon}' = \begin{pmatrix} \varepsilon'_{xx} & \varepsilon'_{xy} & 0 \\ \varepsilon'_{yx} & \varepsilon'_{yy} & 0 \\ 0 & 0 & \varepsilon'_{zz} \end{pmatrix} = \begin{pmatrix} \mu'_{xx} & \mu'_{xy} & 0 \\ \mu'_{yx} & \mu'_{yy} & 0 \\ 0 & 0 & \mu'_{zz} \end{pmatrix} = \bar{\mu}' \quad (7)$$

where $\varepsilon'_{xx} = \left(a_{11}^2 + a_{12}^2\right)/\det(\bar{\Lambda})$, $\varepsilon'_{xy} = (a_{11}a_{21} + a_{12}a_{22})/\det(\bar{\Lambda}) = \varepsilon'_{yx}$, $\varepsilon'_{yy} = \left(a_{21}^2 + a_{22}^2\right)/\det(\bar{\Lambda})$, and $\varepsilon'_{zz} = 1/\det(\bar{\Lambda})$ with $\det(\bar{\Lambda})$ being the determinant of Jacobian matrix.

### 3. Numerical simulations and discussions

Now we perform full-wave simulations in two-dimensional space by a finite element solver (Comsol Multiphysics) to verify the route of transformation media for Gaussian beam collimation. The electromagnetic parameters are assigned in space domain within the region of $0 < x < c$ following EQ. (7). And we only consider transverse magnetic(TM) polarization with magnetic field along z direction.

A Gaussian beam is excited with a beam waist at $x = 0$ as shown with amplitude of magnetic field in Fig.2 (a). We can see that, after propagating through the transformation media, the Gaussian beam is converted into a nearly diffraction-free beam in the region of $x > c$ along $+x$ direction. Compared with the Gaussian beam propagating along $-x$ direction in free-space ($x < 0$), the collimated beam is almost free from divergence as shown in Fig.2 (b) for the normalized power flow.

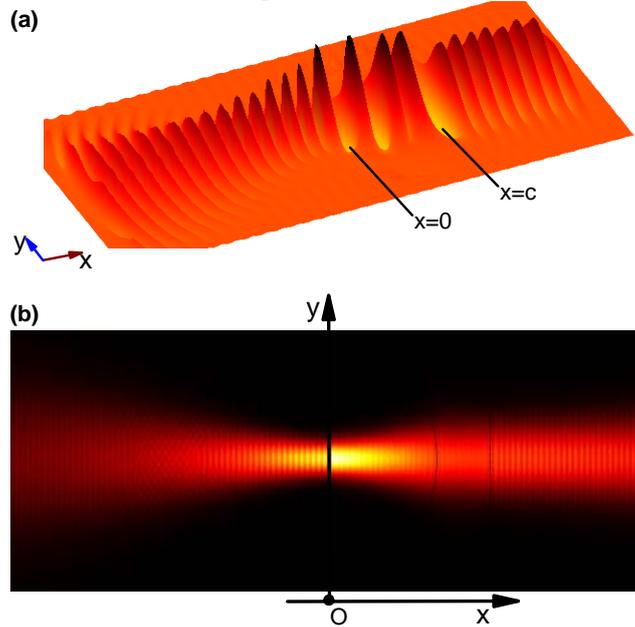

Fig.2 Numerical simulation results of the beam collimation: (a) the spatial distribution of local field $H_z$ at $z = 0$ and (b) energy flux in x-y plane at $x > 0$ and $x < 0$

We can see that the Gaussian beam is converted to a collimated beam within the scale of several wavelengths along both the lateral (*y*) and propagating (*z*) directions, the rectified wave-front at the outgoing interface CC' is very flat with evenly distributed field amplitudes like the one of plane wave. This is exactly a typical example of steering EM wave in a "distorted" EM space created by transformation media. It is noted that the two steps of the transformation we employed play totally different roles. The first step is to shift trapezoidal region $II$ to region $III$ which rectifies the circular wave-fronts of Gaussian beam into planar wave-front at the interface $CC'$ (by EQ.(4)), giving rise to beam collimation; while the second one following EQ.(5) is to ensure the field continuity and impedance matching in the coordinate transformation of extending region $I$, which also leaves us a buffer mechanism in adjusting the waist and other parameters of beam and more freedom controls in adjusting the material parameters of transformation media for practical realization.

Figure 3 presents the spatial distribution of material parameter components. We can see that the components of $\bar{\varepsilon}'$ and $\bar{\mu}'$ are smoothly varying in the space domain with absolute values less than 11.6. The inhomogeneous constitutive parameters can be approximated by discrete transformations [14, 23] and our design is not difficult to be realized by anisotropic metamaterials [24, 25] with gradient-index [14, 26, 27].

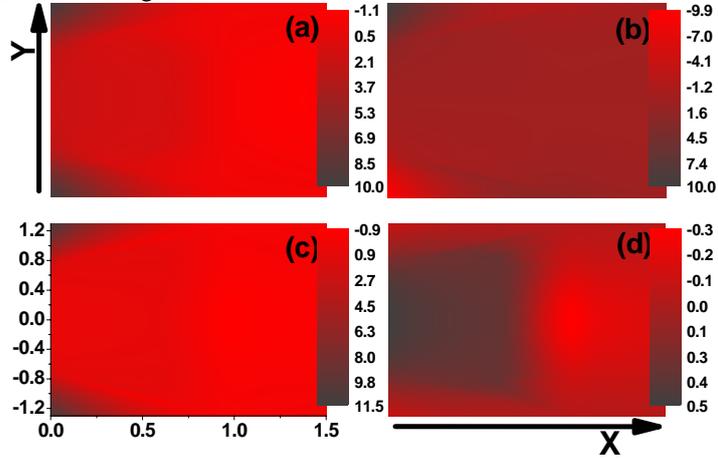

Fig. 3 The spatial distribution in x-y plane at z=0 of the material parameter components: (a) $\varepsilon_{xx} = \mu_{xx}$ (b) $\varepsilon_{xy} = \varepsilon_{yx} = \mu_{yx} = \mu_{xy}$ (c) $\varepsilon_{yy} = \mu_{yy}$ (d) $\varepsilon_{zz} = \mu_{zz}$.

Although we present our work in two-dimensional space to facilitate the discussions and simulations, we also note that it is easy to generalize the transformation to three dimensional case with the rotational symmetry of Gaussian beam.

### 3. Conclusion

In conclusion, we have shown that the collimation of a Gaussian beam can be realized within the scale of several wavelengths, which is far more advantageous than the conventional solution of two-lens confocal system by expanding the beam waist. A two-step transformation scheme is deployed to avoid the singularity of material parameters, facilitating practical implementation. Numerical simulations verify that our transformation successfully converts a Gaussian beam into a plane wave-like beam that is almost free from divergence in propagation.

### Acknowledgements

This work was supported by the National 863 Program of China (Grant No.2006AA03Z407), NSFC (Grant No.10574099, No.60674778), Shanghai Science and Technology Committee and Shanghai Education Committee (No. 06SG24). This work was firstly submitted on October 8, 2009.